\documentclass[prc,reprint,twocolumn,amsmath,amssymb,superscriptaddress]{revtex4-1}
\usepackage{graphicx}
\usepackage{color}

\newcommand{\be}{\begin{equation}}
\newcommand{\ee}{\end{equation}}
\newcommand{\beq}{\begin{eqnarray}}
\newcommand{\eeq}{\end{eqnarray}}
\newcommand{\ba}{\begin{array}}
\newcommand{\ea}{\end{array}}
\newcommand{\crs}{cross section}
\newcommand{\crss}{\crs s}
\newcommand{\gd}{\gamma d}

\newcommand{\gn}{\gamma n}

\newcommand{\gnpip}{\gamma n\!\to\!\pi^-p}

\newcommand{\gdpinn}{\gd\to\pi^+ nn}            
\newcommand{\gdpipp}{\gd\to\pi^- pp}            

\newcommand{\dist}{\displaystyle}
\newcommand{\rpt}{\rule{0pt}{14pt}}
\newcommand{\rptt}{\rule{0pt}{16pt}}
\newcommand{\rttt}{\rule{0pt}{25pt}} 
\newcommand{\ega}{E_{\gamma}}

\newcommand{\bfp}{\mbox {\boldmath $p$}}
\newcommand{\bfq}{\mbox {\boldmath $q$}}
\newcommand{\bfx}{\mbox {\boldmath $x$}}
\newcommand{\bfy}{\mbox {\boldmath $y$}}
\newcommand{\bfk}{\mbox {\boldmath $k$}}
\newcommand{\bfK}{\mbox {\boldmath $K$}}
\newcommand{\bfsig}{\mbox {\boldmath $\sigma$}}
\newcommand{\bfe}{\mbox {\boldmath $e$}}
\newcommand{\bfeps}{\mbox {\boldmath $\epsilon$}}
\newcommand{\bfdeL}{\mbox {\boldmath $\Delta$}}
\newcommand{\bfn}{\mbox {\boldmath $n$}}

\begin{document}

\title{Threshold $\pi^-$ Photoproduction on the Neutron}

\date{\today}

\author{\mbox{W.~J.~Briscoe}}
\affiliation{Institute for Nuclear Studies, Department of Physics, The
        George Washington University, Washington DC 20052, USA}

\author{\mbox{A.~E.~Kudryavtsev}}
\affiliation{National Research Centre ``Kurchatov Institute", Institute
        for Theoretical and Experimental Physics (ITEP), Moscow 117218,
        Russia}
\affiliation{Institute for Nuclear Studies, Department of Physics, The 
	George Washington University, Washington DC 20052, USA}

\author{\mbox{I.~I.~Strakovsky}}
\altaffiliation{Corresponding author; \texttt{igor@gwu.edu}}
\affiliation{Institute for Nuclear Studies, Department of Physics, The 
	George Washington University, Washington DC 20052, USA}

\author{\mbox{V.~E.~Tarasov}}
\affiliation{National Research Centre ``Kurchatov Institute", Institute 
	for Theoretical and Experimental Physics (ITEP), Moscow 117218, 
	Russia}

\author{\mbox{R.~L.~Workman}}
\affiliation{Institute for Nuclear Studies, Department of Physics, The
        George Washington University, Washington DC 20052, USA}

\noaffiliation

\begin{abstract}
Recent data from the PIONS$@$MAX-lab Collaboration, measuring the total 
\crs\ of the pion incoherent photoproduction $\gdpipp$ near threshold, 
have been used to extract the E$_{0+}$ multipole and total 
\crs\ of the reaction $\gnpip$, also near threshold. These are the 
first measurements of the reaction $\gdpipp$ in the threshold region. 
The value of E$_{0+}$ is extracted through a fit to the deuteron data 
in a photoproduction model accounting for final-state interactions.
The model takes an $S$-wave approximation for 
the elementary reaction $\gnpip$ with E$_{0+}\! = $ const in the 
threshold region.  The obtained value E$_{0+} = -31.86\pm 0.8$ 
(in $10^{-3}/m_{\pi^+}$ units) is in agreement with other existing 
results. Model predictions for the total \crs\ $\sigma(\gnpip)$ 
are also given.
\end{abstract}

\maketitle

\clearpage
\section{Introduction}

Pion photoproduction measurements facilitate the understanding of 
the strong force in the low-energy regime. However, most of the experimental 
efforts over the last few decades have focused on neutral pion
production from proton targets $\gamma p\to\pi^0p$~\cite{Ireland:2019uwn}.  
Incoherent 
pion photoproduction on the deuteron is interesting in that it 
provides information 
on the elementary reaction from a neutron target, i.e., $\gamma n\to\pi N$. 
Generally, these latter data are poorly determined due to the paucity of 
neutron reaction data.

A theory of pion photoproduction was constructed in the 1950's.  Kroll 
and Ruderman~\cite{Kroll:1953vq} were the first to derive 
model-independent predictions in the threshold region,a so-called 
Low Energy Theorem (LET), by applying gauge and Lorentz invariance 
to the reaction $\gamma N\to\pi N$. The general formalism for this 
process was developed by Chew and co-workers~\cite{Chew:1957zz} 
(CGLN amplitudes). Vainshtein and Zakharov extended the LET by 
including the hypothesis of a Partially Converted Axial Current 
(PCAC)~\cite{Vainshtein:1972ih}. The derivation of the theorem is 
based on the use of the PCAC hypothesis and on the expansion of 
the amplitudes in powers of $k/m_{int}$ and $q/m_{int}$, where $k$ 
and $q$ are are pion and photon four-momenta and $m_{int}$ is some 
internal mass. This work succeeded in describing the threshold
amplitude as a power series in the ratio $\chi = m_\pi / m$ up to
terms of order $\chi^2$ ($m_{\pi}$ and $m$ are the averaged pion and 
nucleon masses). Somewhat later, Berends and co-workers~\cite{Berends:1967vi}
analysed the existing data in terms of a multipole decomposition 
and extracted the various multipole amplitudes contributing in a
region up to an excitation energy of 500~MeV. These amplitudes are 
vital inputs to low-energy descriptions of hadron physics based 
on the Chiral Perturbation Theory ($\chi$PT)~\cite{Hilt:2013fda}.

Measurements of pion photoproduction on both proton and ``neutron"
targets have a very long history, dating back about 70 years, involving
by the University of Bristol group~\cite{Lattes:1947mw}. 
The first bremsstrahlung facilities produced pioneering results for 
$\gamma p\to\pi^+n$~\cite{McMillan1949} and $\gamma p\to
\pi^0p$~\cite{Panofsky:1950gj}. It is impressive that this work started
two years after the pion discovery in 
1947~\cite{Powell:1947}. The first $\gamma n\to\pi^-p$ 
photoproduction experiment used the 318-MeV photon beam from the 
Berkeley electron synchrotron and a high pressure, low temperature 
deuterium target~\cite{White:1952zz}. Despite all the shortcomings 
of the first measurements (such as large normalization 
uncertainties, wide energy and angular binning, limited angular 
coverage, etc.), those measurements were crucial for the discovery 
of the first baryon resonance, $\Delta$-isobar~\cite{Anderson:1953dgc}.

Present experimental facilities allow some of the most
challenging problems of intermediate energy physics to be studied.
These include the behavior of charged and neutral pion production at
threshold and the electric quadrupole amplitude, E$_0^+$.
Threshold measurements of $\pi^0$ photoproduction, from a
proton target, have been obtained with greater kinematic coverage and 
higher precision than
the associated charged pion photoproduction channels (Table~\ref{tab:tbl1}).
\begin{table}[htb!]
\caption{Threshold energies for pion photoproduction and number
	of measurements at the threshold (below E$_\gamma$ =
	180~MeV) as available in GWU SAID
	database~\protect\cite{SAID}. \label{tab:tbl1}}
\begin{center}
\begin{tabular}{|c|cc|cc|}
\hline
   Reaction         &W$\,$(MeV)&$\!\!$E$\,$(MeV)& $d\sigma/d\Omega$ &$\!\!$Pol \\
\hline
$\gamma p\to\pi^0p$ & 1073.2 &144.7  &1110               & 508 \\
$\gamma n\to\pi^0n$ & 1074.5 &144.7  &   0               &   0 \\
\hline
$\gamma n\to\pi^-p$ & 1077.8 &148.4  &  21               &  12 \\
$\gamma p\to\pi^+n$ & 1079.1 &151.4  & 112               &   0 \\
\hline
\end{tabular}
\end{center}
\end{table}

\begin{figure*}[htb!]
\begin{center}
    \includegraphics[height=3in, keepaspectratio, angle=90]{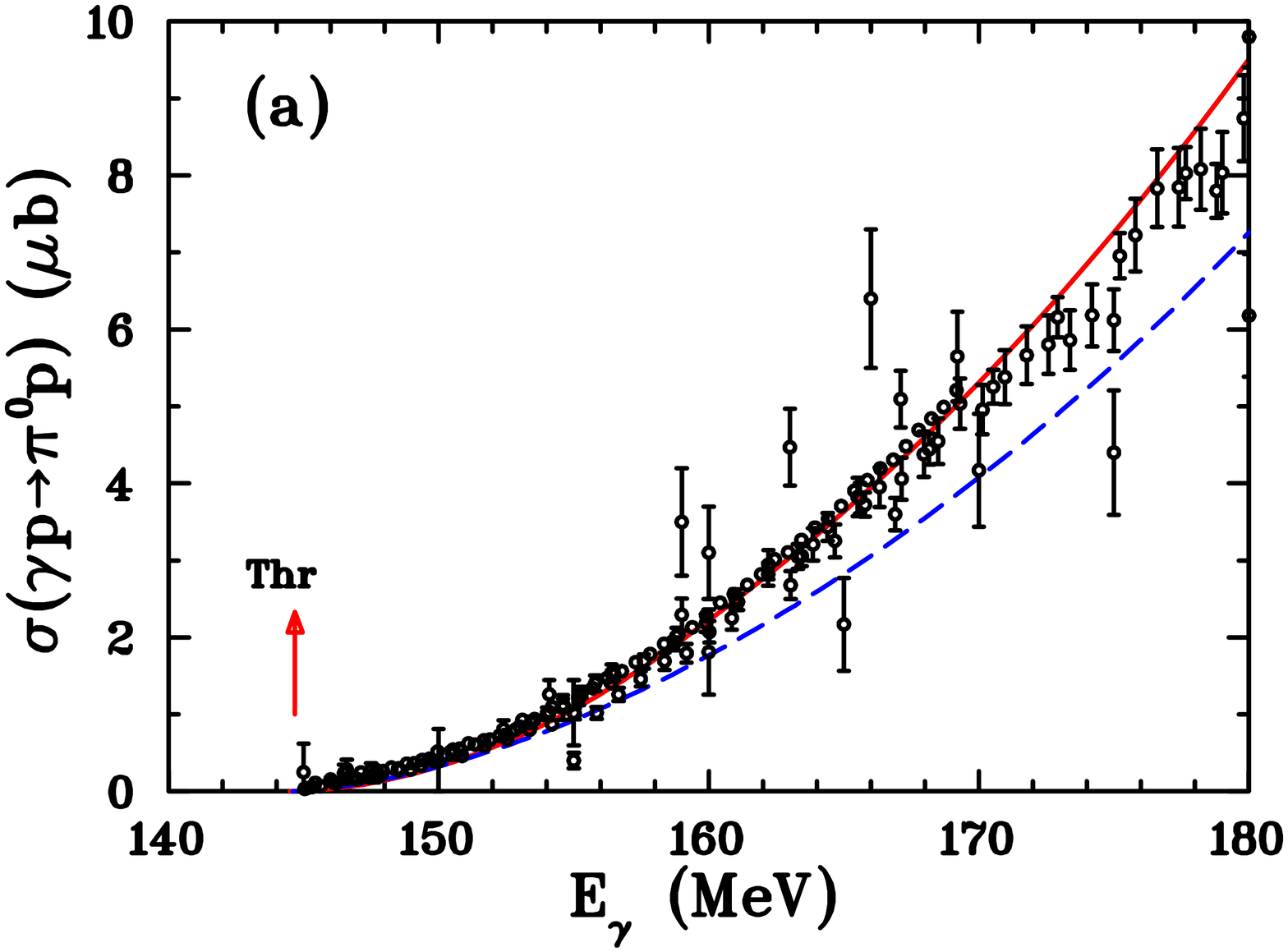}
    \includegraphics[height=3in, keepaspectratio, angle=90]{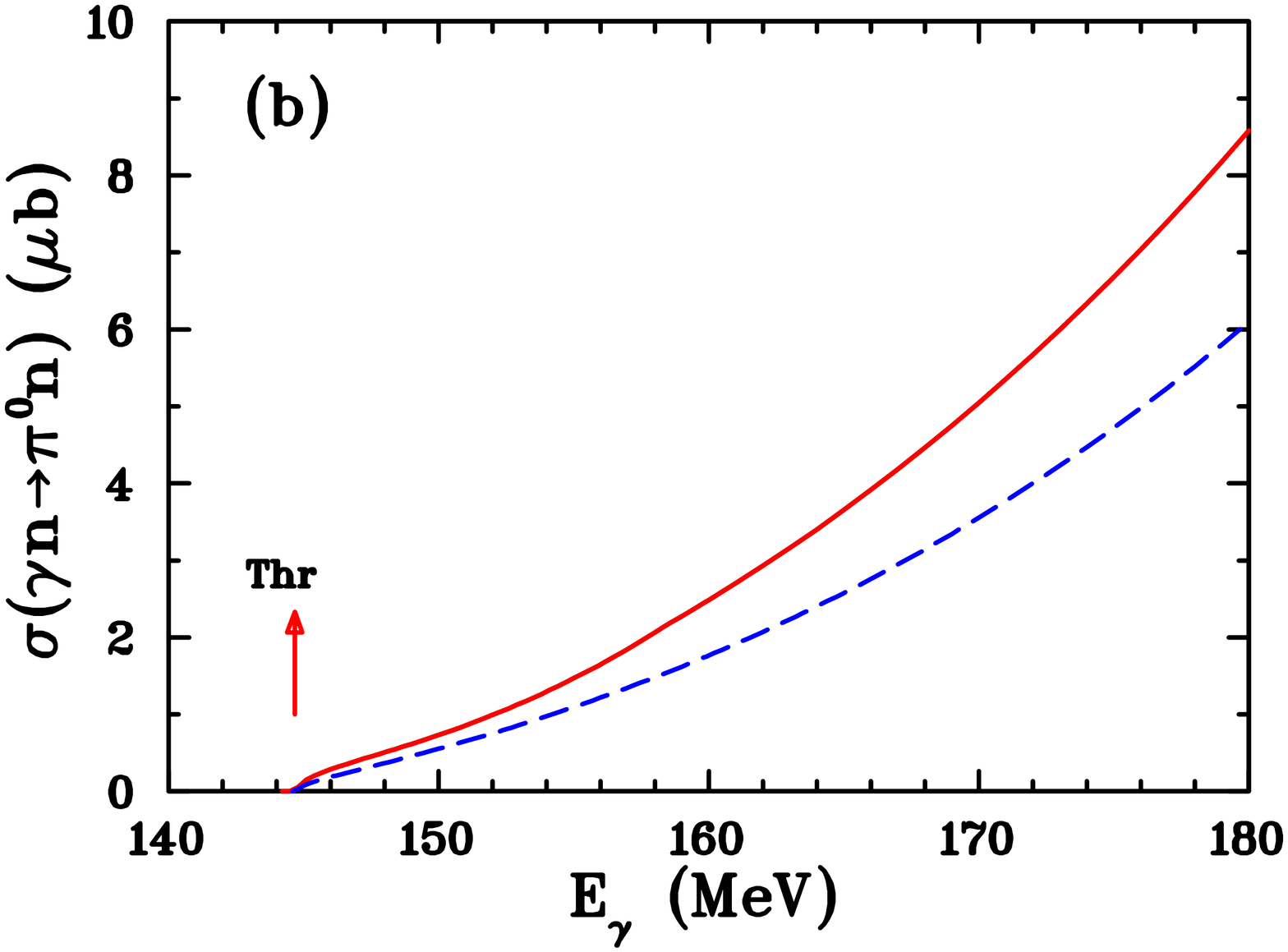}
\end{center}
\begin{center}
    \includegraphics[height=3in, keepaspectratio, angle=90]{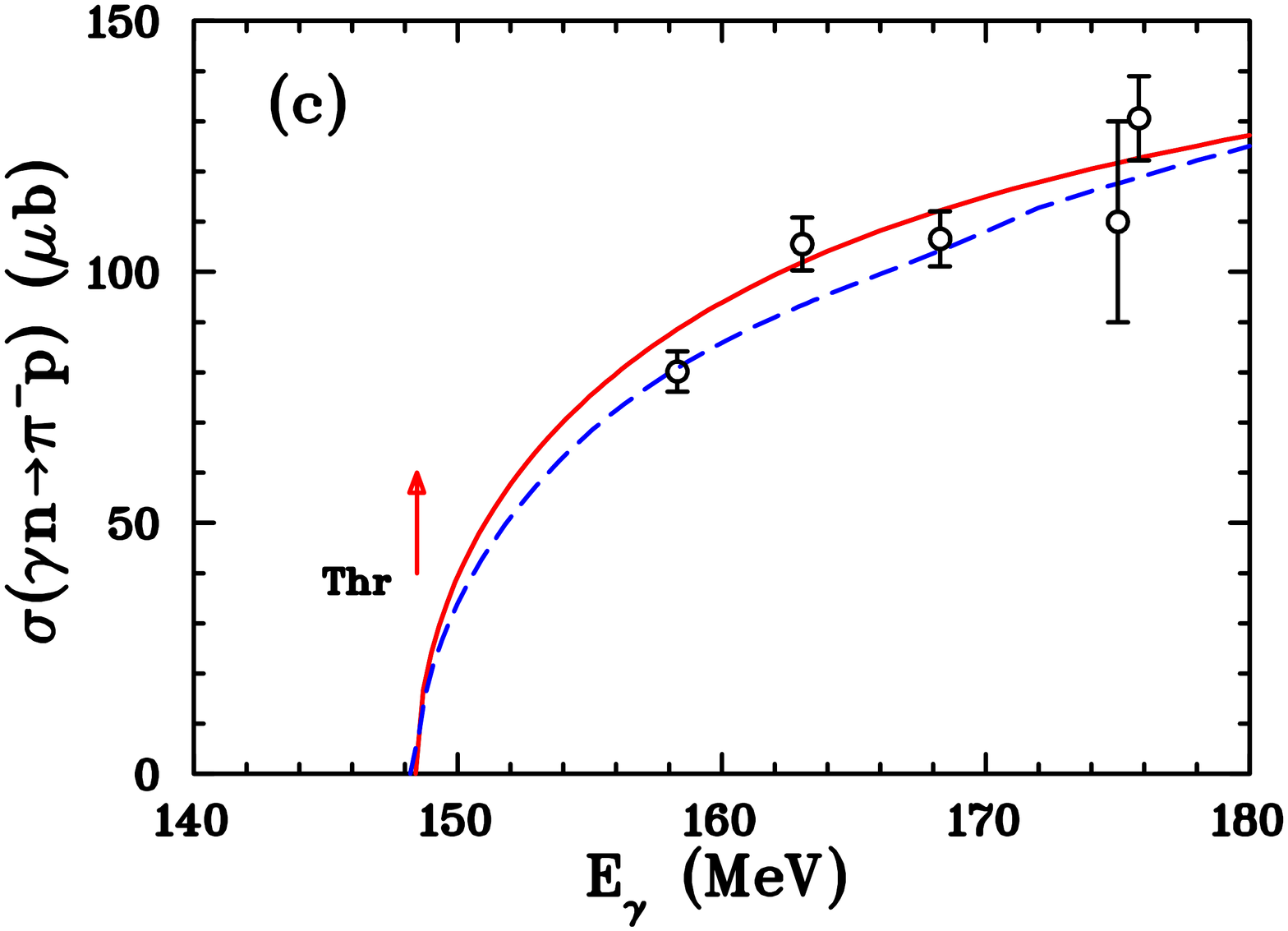}
    \includegraphics[height=3in, keepaspectratio, angle=90]{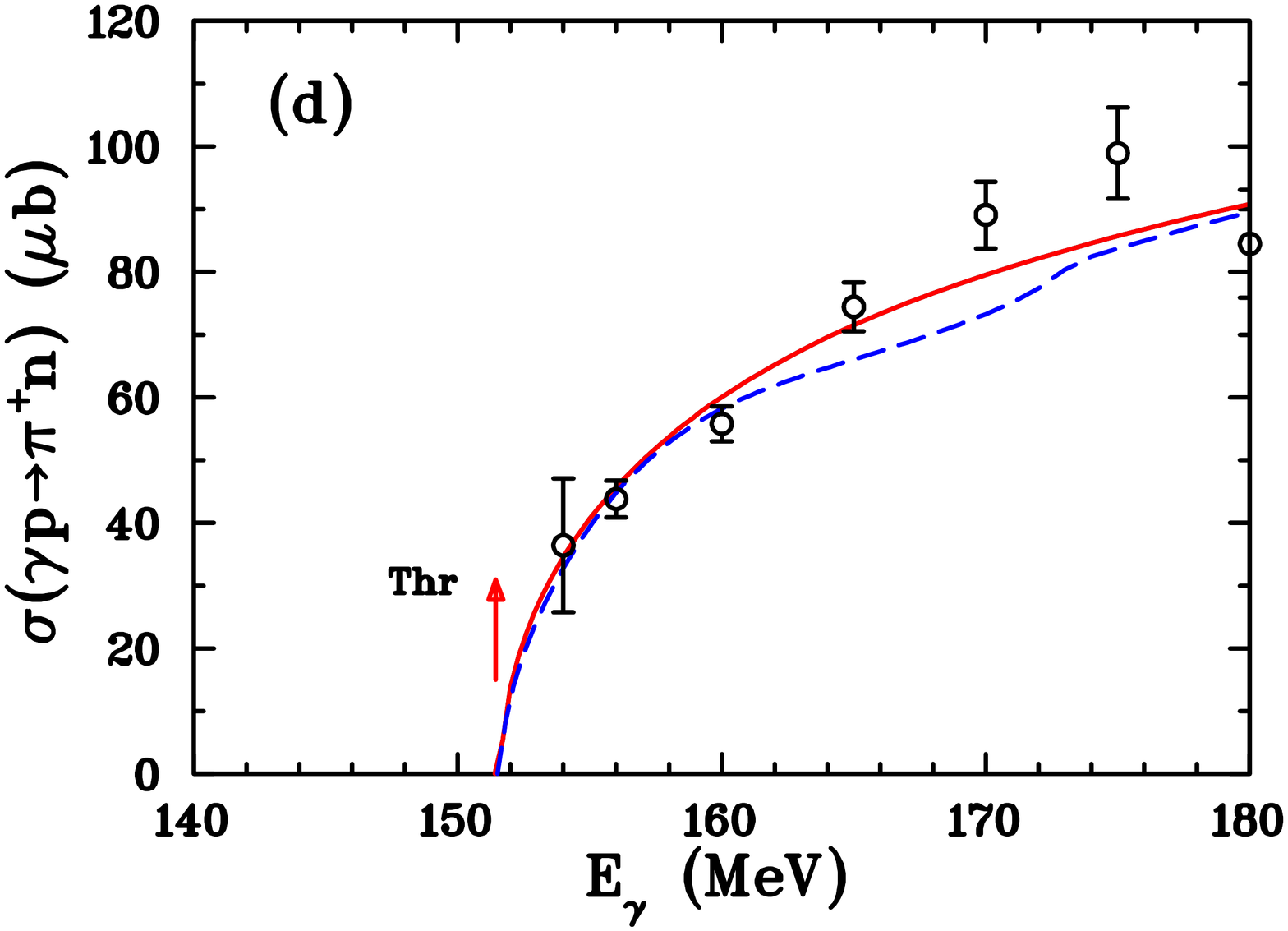}
\end{center}

    \caption{Total cross section of the reactions
        $\gamma p\rightarrow\pi^0p$ (a),
        $\gamma n\rightarrow\pi^0n$ (b),
        $\gamma n\rightarrow\pi^-p$ (c), and
        $\gamma p\rightarrow\pi^+n$ (d).
	Open black circles are the previous
	measurements~\protect\cite{SAID}.
	Plotted uncertainties are statistical and
        systematic in quadrature. Red solid
	(blue dashed) curves are predictions by the
	SAID MA19~\protect\cite{Briscoe:2019cyo}
	(MAID2007~\protect\cite{Drechsel:2007if})
	solution.  Both solutions did not use new
	MAX-lab data in fits.} \label{fig:sgt}
\end{figure*}

The total cross section at the pion production threshold is known for 
$\gamma p\to\pi^0p$ while information about other 
pion photoproduction reactions comes mainly through an extrapolation
of partial-wave analyses (PWA), 
such as SAID~\cite{Briscoe:2019cyo} and \\
MAID~\cite{Drechsel:2007if}, and does not have experimental 
confirmation (Fig.~\ref{fig:sgt}) .

Recently, the PIONS$@$MAX-lab Collaboration has reported total
cross section measurements of the pion incoherent photoproduction
$\gamma d\to\pi^-pp$ at threshold~\cite{Strandberg:2018djk}.
The experiment was performed at the Tagged-Photon
Facility~\cite{Adler:2013} at the MAX~IV Laboratory in Lund, 
Sweden~\cite{Eriksson:2014bta}. Data were collected by three 
very large NaI(Tl) spectrometers BUNI, CATS, and DIANA.  The 
measured total cross section of the reaction $\gamma d\to\pi^-pp$ 
and the comparison with our theoretical predictions was shown in 
Fig.~5 of Ref.~\cite{Strandberg:2018djk}.

The present paper is focused on a determination of the total cross
sections for $\pi^-$ photoproduction on a ``neutron" target,
$\gamma n\to\pi^-p$, utilizing the deuteron measurements,
where model-dependent nuclear 
(final-state interaction) (FSI) corrections play a critical role.

\section{Theoretical Analysis}
\subsection{Extraction of the $\gamma n\to\pi^-p$ Cross Sections}

\begin{figure}
\begin{center}
\includegraphics[height=2.4cm, keepaspectratio]{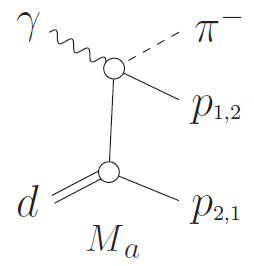}~~~~~~
\includegraphics[height=2.4cm, keepaspectratio]{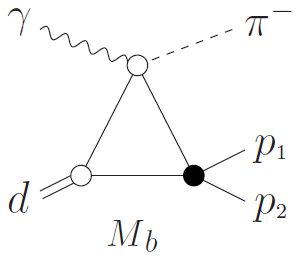}
\end{center}
\begin{center}
\includegraphics[height=2.4cm, keepaspectratio]{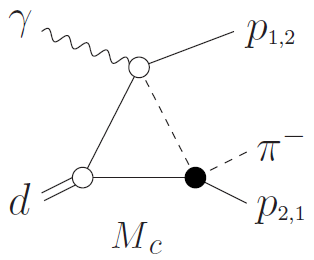}~~~~
\includegraphics[height=2.4cm, keepaspectratio]{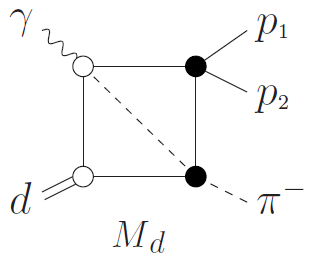}
\end{center}

\vspace{-4mm}
\caption{{\small
     $\!$The~IA$\,(M_a)$,~$NN$-FSI$\,(M_b)$,~$\pi N$-FSI$\,(M_c)$,
     and 2-loop$\,(M_d)$ diagrams for the reaction $\gdpipp$.}}
\label{fig:diag}
\end{figure}

A mathematical description of the FSI model is given in Appendix.
Here the features of this model are summarized. Compared to
the elementary reaction $\gamma n\to\pi^-p$, the 
additional FSI treatment has a non-negligible effect on the cross 
section. The full model~\cite{Tarasov:2011ec} is applied with 
simplifications corresponding to the near-threshold region. 
The four diagrams 
in Fig.~\ref{fig:diag} are calculated, where $M_a$ is the 
Impulse Approximation (IA) term; $M_b$ and $M_c$ are the $N\!N$ 
and $\pi N$ FSI terms; $M_d$ is the $N\!N$-FSI term with pion 
rescattering in the intermediate state (the ``two-loop" term 
added here). Both $M_a$ and $M_c$ are the sums of two terms, 
arising from permutation of the final protons. The total 
amplitude $M_{\gd}$ is taken as the sum $M_{\gd} = M_a + M_b + 
M_c + M_d$.

General expressions for the total \crss\ of the reactions ofinterest can be written as
\be
	\ba{l}\dist
	\sigma(\gnpip)=4\pi\frac{k}{q_{\gn}}\overline{|F_{\gn}|^2},
	\\ \rule{0pt}{18pt}\dist
	\sigma(\gdpipp)=\frac{1}{4q_{\gd}\sqrt{s}}\int\overline
	{|M_{\gd}|^2}\,d\tau_3.
\ea
\label{1}
\ee
Here: $q_{\gn}$ ($k$) and $\overline{|F_{\gn}|^2}$ are the 
center-of-mass (CM) momentum of the initial photon (final pion) 
in the reaction $\gamma n\!\to\!\pi^-p$, and the amplitude 
squared (unpolarized case); $q_{\gd}$ and $\sqrt{s}$ are the 
the CM momentum of the initial photon and total energy in the 
reaction $\gdpipp$; $\overline{|M_{\gd}|^2}$ and $d\tau_3$ are 
the invariant amplitude squared of the reaction $\gdpipp$ 
(unpolarized case) and phase-space element of the final 
$\pi^-pp$ system, where
\be
\ba{l}\dist
	d\tau_3=I\frac{Qp\,dwdzdz_1d\varphi_1}{2\pi(4\pi)^3\sqrt{s}},~~
	p=\sqrt{2\bar\mu w},
	\\ \rptt \dist
	Q=\sqrt{2\bar m(E^\ast-w)},~~~w = M_{pp} - 2m_p.
\ea
\label{2}
\ee
Here: $I = 1/2$ is the symmetry factor for two identical protons; 
$E^\ast = \sqrt{s}\! - \mu - 2m_p$ is the excess energy; 
$\mu$($m_p$) is the $\pi^-$ (proton) mass; $m = (m_p + m_n)/2$; 
$M_{pp}$ is the effective mass of the $pp$ system; $\bar m = 
2m\mu/(2m + \mu)$, $\bar\mu = m\mu/(m + \mu)$; $z\! = 
\cos\theta$, $z_1\! = \cos\theta_1$; $\theta$ is the $\pi^-$ 
polar angle in the reaction rest frame; $\theta_1$ and 
$\varphi_1$ are the polar and azimuthal angles of relative 
motion in the $pp$ system. All the kinematical variables, 
needed to calculate the amplitude $\overline{|M_{\gd}|^2}$, 
can be expressed through $E^\ast$, $w$, $z$, $z_1$, and 
$\varphi_1$.

The ingredients and approximations, used here in the model, are 
as follows.

\vspace{1mm}
1)~In the threshold region, we use the $s$-wave $\gamma n\!\to\!
\pi^-p$ amplitude, given by the $E_{0+}$ multipole, taken to be 
constant. We include only the charged intermediate pion $\pi^-$ 
in the diagrams $M_c$ and $M_d$ since the contribution of 
intermediate $\pi^0$ is suppressed due to a small photoproduction 
$\gamma N\!\to\!\pi^0N$ amplitudes. Thus, $F_{\gn}\! = E_{0+}$ 
and $M_{\gd}\!\sim E_{0+}$. Hereafter $E_{0+}\equiv 
E_{0+}(\gnpip)$. In this approximation, $\sigma(\gnpip)\sim 
E^2_{0+}$ and 
\be
	\sigma(\gdpipp) = E^2_{0+}\,\sigma^{}_0,
\label{3}
\ee
where $\sigma^{}_0$ is $\sigma(\gdpipp)$, calculated according to
Eq.~(\ref{1}) with the factor $E_{0+}$ taken out of the amplitude
$M_{\gd}$, i.e., $\sigma^{}_0$ doesn't depend on $E_{0+}$.

\vspace{1mm}
2)~In the $N\!N$-FSI ($M_b$) and 2-loop ($M_d$) terms, the 
$S$-wave $pp$-scattering amplitude, which also includes the 
Coulomb effects, was taken from Ref.~\cite{Landau:1991wop}. The 
off-shell correction to the $pp$ amplitude was taken into account as 
was done previously, in Refs.~\cite{Levchuk:2006vm,Tarasov:2011ec}, 
by multiplying the on-shell $pp$ amplitude by the monopole 
form factor $F(q,p) = (p^2\! + \!\beta^2)/(q^2\! + \!\beta^2)$.
Here: $q$ and $p$ are the relative momenta of the intermediate 
and final protons; $\beta = 1.2\,$fm.

\vspace{1mm}
3)~In the $\pi\!N$-FSI ($M_c$) and 2-loop ($M_d$) terms, the 
$S$-wave $\pi^-p\,$-scattering amplitude $a_{\pi^-p}\!=b_0-b_1$ is 
used, fixed by the isospin scattering lengths $b_0 = -28$ and $b_1 
= -881$ in $10^{-4}/\mu$ units~\cite{Doring:2004kt}.

\vspace{1mm}
4)~The deuteron wave function (DWF) of the Bonn potential was used 
in parametrization from Ref.~\cite{Machleidt:2000ge}. Both $S$- 
and $D$-wave parts of DWF are included in the IA diagram $M_a$, 
while $D$-wave part is neglected in the diagrams $M_b$, $M_c$, and 
$M_d$.

\vspace{3mm}
The terms $M_{a,b,c,d}$ and the total amplitude squared, 
$\overline{|M_{\gd}|^2}$ (unpolarized case), are written out in 
Appendix (Sections 1 and 2). In the given approximation, the 
integrals over the intermediate states in the loop terms 
$M_{b,c,d}$ are obtained in analytic form (see Appendix, Section 
3).

\begin{figure}
\begin{center}
\includegraphics[height=3.5in, keepaspectratio, angle = 90]{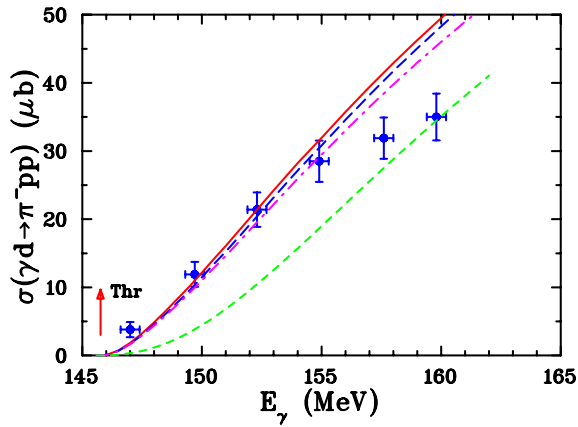}
\end{center}

\vspace{-4mm}
\caption{{\small Total cross section of the reaction $\gdpipp$: 
  blue fulled circles are the MAX-lab 
  data~\protect\cite{Strandberg:2018djk}; 
  E$_\gamma$ is the photon energy in the laboratory frame; the 
  statistical and systematic uncertainties from Table~II of 
  Ref.~\protect\cite{Strandberg:2018djk} are summed in 
  quadrature.
  Green short dashed curve shows the result, obtained with the 
  IA amplitude $M_a$ in Fig.~\protect\ref{fig:diag}. Successive 
  addition of $M_b$($N\!N$-FSI), $M_c$($\pi N$-FSI) and 
  $M_d$(2-loop) terms leads to magenta dash-dotted, blue long
  dashed, and red solid curves, respectively.}} \label{fig:gdeut}
\end{figure}

Now we fit the latest data by the PIONS$@$MAX-lab 
Collaboration~\cite{Strandberg:2018djk} on $\sigma^{exp}(\gdpipp)$ 
close to threshold by the Eq.~(\ref{3}), making use of $E_{0+}$ as 
a free parameter, and obtain $E_{0+}(1\!-\!6) = -31.86\pm 0.8$ 
(in $10^{-3}/\mu$ units). The notation (1$-$6) means that the 
$\chi^2$ fit includes all the 6 data points in Fig.~\ref{fig:gdeut}. 
The curves show the \crss, calculated according to Eq.~(\ref{3}), 
where the red solid one shows the result obtained with the total 
amplitude $M_{\gd}$. The other curves are explained in the figure 
caption. One can see that the main effect of FSI comes from the 
$N\!N$-FSI term $M_b$ (compare magenta dash-dotted, blue long
dashed curves in Fig.~\ref{fig:gdeut}), while the role of the terms 
$M_c$ and $M_d$ is small.

A relatively large disagreement of the model with the data is 
observed close to threshold at E$_\gamma$ = 147~MeV. Excluding 
this ``bad'' 1-st data point from the fit, we obtain $E_{0+}$(2$-$6)$
= -31.75\pm 0.8$ (the same units). Both variants, (1$-$6) and (2$-$6), 
are in agreement with the value $E_{0+} = -31.9$ from 
Ref.~\cite{Drechsel:1992pn}. The model also overestimates the data 
above E$_{\gamma}\!\sim\!156$~MeV. If one excludes two data points 
(5-th and 6-th) at E$_{\gamma} = 157.6$~MeV and 159.8~MeV in 
Fig.~\ref{fig:gdeut}, then the $\chi^2$ fit gives $E_{0+}$(1$-$4)$ =
-33.70\pm 1.2$ and $E_{0+}$(2$-$4)$ = -33.94\pm 1.2$. Suppose this 
discrepancy partly comes from the model approximations with 
energy-independent $\gnpip$ amplitude $E_{0+}\! = \,$const. Let us 
briefly discuss the effects not included here, connected with 
energy dependence of $E_{0+}$ and $P$-wave contribution to the 
$\gnpip$ amplitude. We can roughly estimate these corrections from 
the results of Ref.~\cite{Lensky:2005hb} on the reaction $\gdpinn$ 
in the chiral perturbation theory, where the Born $\gnpip$ 
amplitudes (with a Kroll-Ruderman term) in the threshold region were 
used. At $\Delta\ega\! = \!\ega\!-\!E_{th}\! = \!15$~MeV ($E_{th}$ 
is the threshold energy), the energy-dependent correction to the 
constant $E_{0+}$ decreases the total \crss\ by $\sim 6\,\%$ 
(Fig.~8 of Ref.~\cite{Lensky:2005hb}), while the $P$-wave 
contribution increases it by $\sim 3\,\%$ (Fig.~9 there). 
Approximately the same corrections for the reaction $\gdpipp$ seem 
not enough to improve essentially the discrepancy in 
Fig.~\ref{fig:gdeut} above E$_{\gamma}\!\sim\!156$~MeV. We leave 
these details for a future study.

Table~\ref{tab:tbl2} shows the \crss\ $\sigma(\gnpip)$ from 
Eq.~(\ref{1}) at $F_{\gn}\! = E_{0+}(1\!-\!6) = -31.86\pm 0.8$. 
The results are given at the same values $\Delta\ega\! = \!\ega\! 
- \!E_{th}$ as in Fig.~\ref{fig:gdeut}, i.e., the E$_{\gamma}$'s 
are shifted by the difference $(148.44 - 145.76)\,$MeV) of the 
$\gnpip$ and $\gdpipp$ threshold energies. Total uncertainties 
included statistical and systematical uncertainties of the 
MAX-lab experimental data with the FSI contribution.
\vspace{10mm}
\vspace{10mm}
\begin{figure}
\begin{center}
\includegraphics[height=3.5in, keepaspectratio, angle=90]{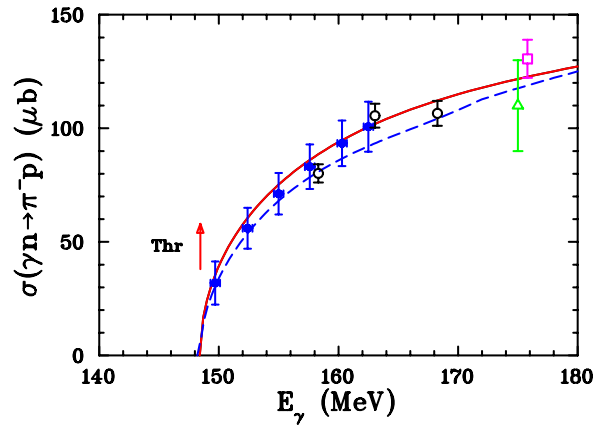}
\end{center}

\vspace{-4mm}
\caption{Total cross section of the reaction
        $\gamma n\rightarrow\pi^-p$.
	Previous measurements for the inverse reaction
	$\pi^-p\to\gamma n$ are from Cornell 
	synchrotron~\protect\cite{White:1960ukk}
	(green open triangle), and 
	TRIUMF~\protect\cite{Salomon:1983xn} (magenta 
	open square) and \protect\cite{Liu:1994} (black 
	open circles). Statistical and systematical 
	uncertainties are summed in quadrature.
	Red solid (blue dashed) curves are predictions 
	by the SAID MA19~\protect\cite{Briscoe:2019cyo}
        (MAID2007~\protect\cite{Drechsel:2007if})
        solution. 
}
\label{fig:gneut}
\end{figure}

As an aside, we have previously used this GW-ITEP FSI code to determine,
at much higher energies, the $\gamma n\to\pi^-p$
differential cross section from $\gamma d\to\pi^-pp$ measurements
with CLAS and A2 at MAMI Collaborations~\cite{Tarasov:2011ec,
Chen:2012yv,Briscoe:2012ni,Mattione:2017fxc,Briscoe:2019cyo}.
In this way, we succeeded in the first determination of neutron
couplings at a set of pole positions~\cite{Mattione:2017fxc,
Briscoe:2019cyo} using these additions to the world data.

%
\begin{table}[t]
\caption{Total cross section for $\pi^-$ photoproduction
	on the neutron with statistical and systematic
	uncertainties in quadratures.
}
\centering
\begin{tabular}{|c|c|ccc|}
\hline
                 &               & Exp  & Exp  & $E_{0+}$ fit \\
  $E_\gamma$     &    $\sigma$   & Stat & Sys  & Sys \\
     (MeV)       &    ($\mu$b)   & (\%) & (\%) & (\%)\\
\hline
  149.7$\pm$ 0.4 & 31.9$\pm$ 9.5 & 5.3  & 28.9 & 4.9 \\
  152.4$\pm$ 0.4 & 56.0$\pm$ 9.0 & 2.5  & 15.1 & 4.9 \\
  155.0$\pm$ 0.4 & 71.2$\pm$ 9.1 & 1.4  & 11.7 & 4.9 \\
  157.6$\pm$ 0.4 & 83.1$\pm$ 9.8 & 1.8  & 10.5 & 4.9 \\
  160.3$\pm$ 0.4 & 93.4$\pm$10.0 & 1.3  &  9.4 & 4.9 \\
  162.5$\pm$ 0.4 &100.7$\pm$11.0 & 1.4  &  9.7 & 4.9 \\
\hline
\end{tabular}
\label{tab:tbl2}
\end{table}

\subsection{Summary and Impact of new MAX-lab data for Partial-Wave Analysis}

In summary, total cross sections for the $\gamma n\to\pi^-p$ have
been taken at photon energies within 1-2~MeV of the reaction
threshold. These data are in good agreement with predictions from
previous analyses, such as SAID and MAID. 

In general, to prevent double counting, we do not use total cross section
data (integral over differential cross sections) in the SAID partial-wave analyses. 
analysis.  Specifically at the threshold, it is hard to cover a
full angular range and some assumptions are required
to determine a total cross section. 

The importance of improving the $\gamma$n database relative to the
$\gamma$p database is directly related to the fact that the
electromagnetic interaction does not conserve isospin symmetry.
The amplitude for the reactions $\gamma N\to\pi N$ factors into
distinct $I = 1/2$ and $I = 3/2$ isospin components,
$A_{\gamma,\pi^\pm} = \sqrt{2}(A_{p/n}^{I = 1/2} \mp A^{I = 
3/2})$ (see Ref.~\cite{Drechsel:1992pn}).  This expression 
indicates that the $I = 3/2$ multipoles can be entirely determined 
from proton target data. However, measurements from datasets with 
both neutron and proton targets are required to determine the 
isospin $I = 1/2$ amplitudes.

\section{Acknowledgements}

This work was supported in part by the U.S. Department of Energy, 
Office of Science, Office of Nuclear Physics under Awards No. 
DE--SC0016583 and DE--SC0016582. The authors A.E.K. and V.E.T. 
acknowledge the support of the RFBR under Award No. 16--02--00767.

\vspace{3mm}
\centerline{\bf APPENDIX:}
\centerline{\bf THE REACTION AMPLITUDE}
\vspace{2mm}

{\centerline{\bf 1.~The Reaction Amplitude}
\vspace{1mm}

The invariant amplitude $M_{\gd}$ of the reaction $\gdpipp$ can be
written as
$$
\ba{c}
       M_{\gd} = c\varphi^ + _1(L\!+i\bfK\cdot\bfsig)\varphi^c_2,
\\ \rpt
       c = 16\pi W\!\sqrt{m}~~(W\! = \!m+\!\mu),
\\ \rpt
       L = L_a\! + L_b\! + L_c\! + L_d, 
\\ \rpt
       L_a\! = L^{(s)}_a\! + L^{(d)}_a,
\\ \rpt
   \bfK\! = \!\bfK_a\! + \!\bfK_b\! + \!\bfK_c\! + \!\bfK_d,
\\ \rpt
       \bfK_a\! = \bfK^{(s)}_a\! + \bfK^{(d)}_a.
\ea
\eqno{(A.1)}
$$
Here: $\varphi_{1,2}$ are the spinors of the final protons
($\varphi^+\varphi\equiv 1$) and $\varphi^c\equiv\sigma_2\varphi^\ast$;
the subscripts $a,b,c,d$ (in $L_a$,.., $\bfK_a$,..) correspond to the
diagrams in Fig.~\ref{fig:diag};
$L^{(s)}_a\!$ and $\bfK^{(s)}_a$ ($L^{(d)}_a\!$ and $\bfK^{(d)}_a$) 
are the IA amplitudes with $s$-wave ($d$-wave) part of the DWF. The 
amplitudes $L_a$,.. and $\bfK_a$,.. are given below, where $\bfe$ and 
$\bfeps$ are the photon and deuteron polarization three-vectors, 
respectively. Hereafter: $\bfq$, $\bfk$, $\bfp_{1,2}$ stand for the 
three-momenta of the initial photon, final pion and final protons, 
respectively, in the laboratory frame.

{a)~\underline{IA Terms}:
$$
\ba{l}
     ~~L^{(s)}_a = x_aE_{0+}(\bfe\cdot\bfeps),~~ x_a\! = \!f_1\! + \!f_2,
\\ \rpt
     ~~L^{(d)}_a = -\Bigl[g_1(\bfe\cdot\bfn_1)(\bfe\cdot\bfn_2) +
\\ \rpt
     ~~~~~~~~~~~+g_2(\bfe\cdot\bfn_2)(\bfeps\cdot\bfn_1)\Bigr]E_{0+},
\\ \rpt
     ~~\bfK^{(s)}_a = y_aE_{0+}[\bfe\times\!\bfeps],~~ y_a\! = \!f_2\!-\!f_1,
\\ \rpt
     ~~\bfK^{(d)}_a = \Bigl(g_2(\bfeps\cdot\bfn_2)[\bfn_2\!\times\!\bfe] -
\\ \rpt
     ~~~~~~~~~~~-g_1(\bfeps\cdot\bfn_1)[\bfn_1\!\!\times\!\bfe]\Bigr)E_{0+};
\\ \rule{0pt}{22pt}\dist
    ~~~~f_{1,2}\! = \frac{u(p_{1,2})}{\sqrt{2}} + \frac{w(p_{1,2})}{2},
\\ \rptt\dist
    ~~~~g_{1,2}\! = \frac{3}{2}\,w(p_{1,2}),
         ~~~\bfn_{1,2}\! = \frac{\bfp_{1,2}}{p_{1,2}}.
\ea
\eqno{(A.2)}
$$
Here: $\bfn_{1,2}$ -- the unit vectors; $u(p)$ and $w(p)$ are the $S$- 
and $D$-wave parts of the DWF. We use DWF~\cite{Machleidt:2000ge}, 
parametrized in the form
$$
        u(p) = \sum_j\frac{C_j}{p^2\! + m^2_j},~~
        w(p) = \sum_j\frac{D_j}{p^2\! + m^2_j}~
\eqno{(A.3)}
$$
with normalization $\int\!d\bfp\,[u^2(p)\! + w^2(p)] = (2\pi)^3$.

\vspace{2mm}
{b)~\underline{$pp$-FSI Terms}:
$$
\ba{c}
	L_b = x_b E_{0+}(\bfe\cdot\bfeps),~~ \bfK_b = 0,
\\ \rpt
	x_b = 2 I_{pp}(p,\beta,\Delta)f_{pp}(p),
\\ \rttt\dist
	I_{pp}(p,\beta,\Delta) = \int\!\frac{d\bfx\,f(x,p)u(|\bfx\! 
	+ \!\bfdeL|)} {2\pi^2\sqrt{2}\,(x^2\! - \!p^2\!-i0)},
\\ \rttt\dist
	f(x,p) = \frac{p^2\! + \!\beta^2}{x^2\! + \!\beta^2},
	~~\bfdeL = \frac{1}{2}(\bfp_1\! + \!\bfp_2).
\ea
\eqno{(A.4)}
$$
Here: $f_{pp}(p)$ is the on-shell $S$-wave $pp$-scattering amplitude
in the Effective-Range-Approximation with \\ Coulomb effects
included~\cite{Landau:1991wop}; $\bfx$ is the relative three-momentum 
of the intermediate nucleons; $f(x,p)$ is the formfactor in the 
off-shell $pp$-scattering amplitude $f^{of\!f}_{pp}(x,p) = 
f(x,p)f_{pp}(p)$ with parameter $\beta = 1.2\,$fm, used 
earlier~\cite{Tarasov:2011ec,Levchuk:2006vm}. 
The integral $I_{pp}(p,\beta,\Delta)$ is written out in Eqs.~(A.9) 
and (A.10).

\vspace{2mm}
{c)~\underline{$\pi N$-FSI Terms}:
$$
\ba{c}
     L_c\! = x_cE_{0+}a_{\pi^-\!p}\,(\bfe\cdot\bfeps),
   ~~~x_c\! = I_1\! + \!I_2; 
\\ \rptt\dist
    \bfK_c\! = y_cE_{0+}a_{\pi^-\!p}\,[\bfe\!\times\!\bfeps],
   ~~y_c\!=I_1\!-\!I_2;
\\ \rptt\dist
    I_i\! = \!I(k^2_i,\Delta_i),
  ~~~\bfdeL_i\! = \!\frac{m}{m\! + \!\mu}(\bfk\! + \!\bfp_i).
\ea
\eqno{(A.5)}
$$
Here: $k_i$ are the relative momenta in the pion-proton pairs
$\pi^-p_i\,(i\! = \!1,\!2)$; $a_{\pi^-\!p}$ is the $\pi^-p$-scattering
amplitude in the scattering-length approximation (see the main text).
The integral $I(k^2_{1,2},\Delta_{1,2})$ is written out below in 
Eq.~(A.9).

\vspace{2mm}
{d)~\underline{2-loop Terms}:
$$
\ba{c}
        L_d = x_d E_{0+}(\bfe\cdot\bfeps),~~ \bfK_d=0,
\\ \rpt\dist
        x_d = 2K(p,b,\Delta)f_{pp}(p)a_{\pi^-\!p},
\\ \rule{0pt}{20pt}\dist
        K(p,b,\Delta) = \frac{m\! + \!\mu\!}{m}\,\times
\\ \rule{0pt}{22pt}\dist
         \times\!\int\!\!
        \frac{d\bfx d\bfy\,u(|\bfx\! + \!\bfy\!-\!\Delta|)f(x,p)}
        {4\pi^4\sqrt{2}\,(x^2\! - \!p^2\!-i0)(y^2\! - \!b^2\! - i0)},
\\ \rule{0pt}{18pt}\dist
        \bfdeL\! = \frac{1}{2}(\bfq\! + \!\bfk),~~
        b^2\! = \!2\mu(\sqrt{s} - \sqrt{s_0})\ge 0.
\ea
\eqno{(A.6)}
$$
Here: $\sqrt{s_0}\! = 2m_p\! + \!\mu$; $f(x,p)$ is given in 
Eq.~(A.4); the denominator $(y^2\! - \!b^2\! - \!i0)$ of the 
pion propagator is obtained, neglecting the kinetic energies
(static approximation) of the intermediate nucleons. The 
expression for $K(p,b,\Delta)$ is given in Eqs.~(A.11) and 
(A.12).

\vspace{4mm}
{\centerline{\bf 2.~The Square of the Amplitude}
\vspace{1mm}

The square of the amplitude~(A.1) for unpolarized nucleons
is $|M_{\gd}|^2\! = 2c^2\,(|L|^2\! + |\bfK|^2)$. Averaging 
it over the photon and deuteron polarization states, we write
$$
\ba{c}
   \overline{|M_{\gd}|^2} = 2c^2\,(\overline{|L|^2}\! + 
   \!\overline{|\bfK|^2}).
\ea
\eqno{(A.7)}
$$
Making use of Eqs.~(A.2),(A.4),(A.5), and (A.6), we have
$$
\ba{l}\dist
    L = A E_{0+}(\bfe\cdot\bfeps) + \!L^{(d)}_a,
 ~~ A\! = x_a\! + \!x_b\!+\!x_c\! + \!x_d,
\\ \rpt\dist
   \bfK\! = B E_{0+}[\bfe\!\times\bfeps] + \!\bfK^{(d)}_a,
	~~B\! = y_a\! + y_c.
\ea
$$
Then, we obtain
$$
\ba{c}\dist
        \overline{|L|^2} = \frac{1}{3}\biggl[|A|^2\!
        - (g_1 n^2_{1t}\! + g_2 n^2_{2t}) {\rm Re}[A]+~~~~
\\ \rptt\dist
        + \frac{1}{2}\,(g^2_1 n^2_{1t}\! + g^2_2 n^2_{2t})
\\ \rptt\dist
        + g_1g_2(\bfn_1\cdot\bfn_2)(\bfn_{1t}\cdot\bfn_{2t})\biggr] 
	E^2_{0+},
\\ \rttt\dist
        \overline{|\bfK|^2} = \frac{1}{3}\biggl[2|B|^2 +
\\ \rule{0pt}{16pt}\dist
       + [g_1(1\!+n^2_{1z})\! - g_2(1\! + n^2_{2z})] {\rm Re}[B]+
\\ \rule{0pt}{20pt}\dist
       + \frac{1}{2}\,[g^2_1(1\! + n^2_{1z})\!+g^2_2(1\! 
       + n^2_{2z})]-
\\ \rule{0pt}{18pt}\dist
       - g_1g_2(\bfn_1\cdot\bfn_2)[(\bfn_1\cdot\bfn_2) + 
n_{1z}n_{2z}]\biggr] 
	E^2_{0+}.
\ea
\eqno{(A.8)}
$$
Here: $\bfn_{1t,2t}$ and $n_{1z,2z}$ are, respectively, the 
transverse parts and $z$-components of the unit vectors 
$\bfn_{1,2}$, defined in Eqs.~(A.2), with $z$-axis along the 
photon three-momentum $\bfq$ in the laboratory frame.

\vspace{2mm}
{\centerline{\bf 3.~The Integrals}
\vspace{1mm}

The integral $I_{pp}(p,\beta,\Delta)$ in Eqs.(A.4) can be 
rewritten as
$$
\ba{c}
        I_{pp}(p,\beta,\Delta) = I(p^2,\Delta) - I(-\beta^2,\Delta),
\\ \rule{0pt}{22pt}\dist
        I(a^2\!,\Delta) = \int\!\frac{d\bfx\,u(|\bfx\! + \!\bfdeL|)}
        {2\pi^2\sqrt{2}\,(x^2\! - \!a^2\!-i0)}.
\ea
\eqno{(A.9)}
$$
For the DWF, given in the form~(A.3), we obtain

$$
\ba{l}\dist
        I(a^2\!>\!0,\Delta) =
        \!\sum_j\!\frac{C_j}{2\Delta\sqrt{2}}
        \biggl[\arctan\frac{|a|\! + \!\Delta}{m_j}-
\\ \rule{0pt}{20pt}\dist
        - \arctan\frac{|a|\! - \!\Delta}{m_j}
        + \frac{i}{2}\ln\frac{m^2_j\! + \!(|a|\! 
	+ \!\Delta)^2}{m^2_j\! 
	+ \!(|a|\!-\!\Delta)^2}
\biggr],
\\ \rule{0pt}{22pt}\dist
        I(a^2\!<\!0,\Delta) =
        \!\sum_j\frac{C_j}{\Delta\sqrt{2}}
        \,\arctan\frac{\Delta}{m_j\! + |a|}.
\ea
\eqno{(A.10)}
$$
The integral $K(p,b,\Delta)$ in Eqs.(A.6) can be written as
$$
\ba{c}
     K(p,b,\Delta) = K_0(p^2\!,b^2\!,\Delta) -
     \!K_0(-\beta^2\!,b^2\!,\Delta),
\\ \rule{0pt}{20pt}\dist
     K_0(a^2\!,b^2\!,\Delta) = \frac{m\! + \!\mu}{m}\times
\\ \rule{0pt}{20pt}\dist
     \times\int\!
     \frac{d\bfx d\bfy\,u(|\bfx\! + \!\bfy\! - \!\Delta|)}
     {4\pi^4\sqrt{2}\,(x^2\! - \!a^2\! - i0)(y^2\! - \!b^2\!-i0)},
\ea
\eqno{(A.11)}
$$
For the DWF of the type~(A.3), we obtain
$$
\ba{c}\dist
      K_0(a^2\!,b^2\!,\Delta) =
     \!\sum_j\frac{C_j}{\Delta\sqrt{2}}\Bigl[U_j(a^2\!,b^2\!,\Delta) -
\\ \dist
~~~~~~~~~~~~~~~~~~-U_j(a^2\!,b^2\!, - \Delta)\Bigr], 
\\ \rpt\dist
        U_j(a^2\!,b^2\!,\Delta) = -x_jA_j\! - yL_j+
\\ \rpt\dist
~~~~~~~~~~~~~~~~~~+i(yA_j\! - x_jL_j),~
\\ \rptt\dist
        L_j\! = \frac{1}{2}\ln(x^2_j\! + \!y^2),~ A_j\! 
	= \arctan\frac{y}{x_j};
\\ \rpt\dist
        a^2\!>0\!:~ x_j\! = m_j,~ y = \!|a| + \!|b| + \Delta;
\\ \dist
        a^2\!<0\!:~ x_j\! = m_j\! + \!|a|,~~ y = \!|b| + \Delta.
\ea
\eqno{(A.12)}
$$


\begin{thebibliography}{99}
\bibitem{Ireland:2019uwn}
  D.~G.~Ireland, E.~Pasyuk, and I.~Strakovsky,
  Prog.\ Part.\ Nucl.\ Phys.\  {\bf 111}, 103752 (2020).
\bibitem{Kroll:1953vq}
  N.~M.~Kroll and M.~A.~Ruderman,
  Phys.\ Rev.\ {\bf 93}, 233 (1954).
\bibitem{Chew:1957zz}
  G.~F.~Chew, M.~L.~Goldberger, F.~E.~Low, and Y.~Nambu,
  Phys.\ Rev.\  {\bf 106}, 1337 (1957).
\bibitem{Vainshtein:1972ih}
  A.~I.~Vainshtein and V.~I.~Zakharov,
  Nucl.\ Phys.\ B {\bf 36}, 589 (1972).
\bibitem{Berends:1967vi} 
  F.~A.~Berends, A.~Donnachie, and D.~L.~Weaver,
  Nucl.\ Phys.\ B {\bf 4}, 1 (1967).
\bibitem{Hilt:2013fda} 
  M.~Hilt, B.~C.~Lehnhart, S.~Scherer, and L.~Tiator,
  Phys.\ Rev.\ C {\bf 88}, 055207 (2013).
\bibitem{Lattes:1947mw}
  C.~M.~G.~Lattes, H.~Muirhead, G.~P.~S.~Occhialini, and C.~F.~Powell,
  Nature {\bf 159}, 694 (1947).
\bibitem{McMillan1949}
  E.~M.~McMillan, J.~M.~Peterson, and R.~S.~White, Science, \textbf{110}, 579 (1949).
\bibitem{Panofsky:1950gj}
  J.~Steinberger, W.~K.~H.~Panofsky, and J.~Steller,
  Phys.\ Rev.\  {\bf 78}, 802 (1950).
\bibitem{Powell:1947} C.~M.~G.~Lattes, H.~Muirhead, G.~P.~S.~Occhialini, and
  C.~F.~Powell, Nature, \textbf{159}, 694 (1947).
\bibitem{White:1952zz}
  R.~S.~White, M.~J.~Jacobson, and A.~G.~Schulz,
  Phys.\ Rev.\  {\bf 88}, 836 (1952).
\bibitem{Anderson:1953dgc}
  H.~L.~Anderson, E.~Fermi, R.~Martin, and D.~E.~Nagle,
  Phys.\ Rev.\  {\bf 91}, no. 1, 155 (1953).
\bibitem{SAID} W.~J.~Briscoe, M.~D\"oring, H.~Haberzettl, I.~I.~Strakovsky, 
  and R.~L.~Workman,
  Institute of Nuclear Studies of The George Washington University Database;
  http://gwdac.phys.gwu.edu/ .
\bibitem{Briscoe:2019cyo}
  W.~J.~Briscoe {\it et al.} [A2 Collaboration],
  Phys.\ Rev.\ C {\bf 100}, no. 6, 065205 (2019).
\bibitem{Drechsel:2007if}
  D.~Drechsel, S.~S.~Kamalov, and L.~Tiator,
  Eur.\ Phys.\ J.\ A {\bf 34}, 69 (2007).
\bibitem{Strandberg:2018djk}
  B.~Strandberg {\it et al.} [PIONS$@$MAX-lab Collaboration],
  Phys.\ Rev.\ C {\bf 101}, no. 3, 035207 (2020).
\bibitem{Adler:2013}
  J.-O.~Adler \textit{et al.},
  Nucl.\ Instrum.\ Methods\ Phys.\ Res.\ Sect.\ A\ \textbf{715}, 1 (2013).
\bibitem{Eriksson:2014bta}
  M.~Eriksson,
  in: 5th International Particle Accelerator Conference (IPAC 2014)
  Proceedings, Editors: Ch. Petit-Jean-Genaz \textit{et al.}
  Jun 2014. Dresden, Germany.
\bibitem{Tarasov:2011ec}
  V.~E.~Tarasov, W.~J.~Briscoe, H.~Gao, A.~E.~Kudryavtsev, and I.~I.~Strakovsky,
  Phys.\ Rev.\ C {\bf 84}, 035203 (2011).
\bibitem{Landau:1991wop}
  L.~D.~Landau and E.~M.~Lifshits,
  ``Quantum Mechanics : Non-Relativistic Theory,''
  (Butterworth-Heinemann, 1977).
\bibitem{Levchuk:2006vm}
  M.~I.~Levchuk, A.~Y.~Loginov, A.~A.~Sidorov, V.~N.~Stibunov, and M.~Schumacher,
  Phys.\ Rev.\ C {\bf 74}, 014004 (2006).
\bibitem{Doring:2004kt}
  M.~Doring, E.~Oset, and M.~J.~Vicente Vacas,
  Phys.\ Rev.\ C {\bf 70}, 045203 (2004).
\bibitem{Machleidt:2000ge}
  R.~Machleidt,
  Phys.\ Rev.\ C {\bf 63}, 024001 (2001).
\bibitem{Drechsel:1992pn}
  D.~Drechsel and L.~Tiator,
  J.\ Phys.\ G {\bf 18}, 449 (1992).
\bibitem{Lensky:2005hb}
  V.~Lensky, V.~Baru, J.~Haidenbauer, C.~Hanhart, A.~E.~Kudryavtsev, and
  U.-G.~Mei{\ss}ner,
  Eur.\ Phys.\ J.\ A {\bf 26}, 107 (2005).
\bibitem{Chen:2012yv} 
  W.~Chen {\it et al.},
  Phys.\ Rev.\ C {\bf 86}, 015206 (2012).
\bibitem{Briscoe:2012ni} 
  W.~J.~Briscoe, A.~E.~Kudryavtsev, P.~Pedroni, I.~I.~Strakovsky, V.~E.~Tarasov, and R.~L.~Workman,
  Phys.\ Rev.\ C {\bf 86}, 065207 (2012).
\bibitem{Mattione:2017fxc} 
  P.~T.~Mattione {\it et al.} [CLAS Collaboration],
  Phys.\ Rev.\ C {\bf 96}, no. 3, 035204 (2017).
\bibitem{Bernard:1996ti}
  V.~Bernard, N.~Kaiser, and U.~G.~Mei{\ss}ner,
  Phys.\ Lett.\ B {\bf 383}, 116 (1996).
\bibitem{White:1960ukk} 
  D.~H.~White, R.~M.~Schectman, and B.~M.~Chasan,
  Phys.\ Rev.\  {\bf 120}, no. 2, 614 (1960).
\bibitem{Salomon:1983xn} 
  M.~Salomon, D.~F.~Measday, J.~M.~Poutissou, and B.~C.~Robertson,
  Nucl.\ Phys.\ A {\bf 414}, 493 (1984).
\bibitem{Liu:1994}
  K.~Liu, Ph.~D.~Thesis, University of Kentucky, 1994.
\end{thebibliography}
\end{document}